\documentclass[reprint,onecolumn,amsmath,amssymb]{revtex4}
\usepackage{graphicx}
\begin{document}

\title{Induced long-time correlations in a two-component lattice gas}

\author{Oleksandr~V.~Kliushnychenko}
\author{Sergei~P.~Lukyanets}
\email[Email address:]{lukyan@iop.kiev.ua}
\affiliation{Institute of Physics, National Academy of Sciences, Prospect Nauky 46, 03028 Kyiv, Ukraine}

\begin{abstract}
The distinguishability of at least two species of particles in the classical lattice gas with no interactions except hard-core exclusion entails additional interparticle correlations. A nonlinear mixing flow appears and manifests itself most pronounced in the case of significant difference between mobilities of species.
It may result in the induced correlations for the slow component mediated by the fast one.
In the quasi-one-dimensional case, the long-time correlations are demonstrated to take place in the slow component, that is similar to the hydrodynamic correlations between colloidal particles.
In the adiabatic approximation, these correlations may come into play only in the non-equilibrium case with the flow of the fast component present in the system.
\end{abstract}

\maketitle

\section{\label{intro}Introduction}
Multi\-component systems often exhibit unusual behavior and transport effects \cite{tahir-kheli_correlated_1983,schmittmann_statistical_1995,leung_drifting_1997,hipolito_effects_2003,lukyanets_drift_2010}. Many of these effects are caused by the additional correlations which are associated with particle distinguish\-ability. The short-range ``back correlations'' (memory effect) \cite{tahir-kheli_correlated_1983} in the classical two-component lattice gas with excluded volume constraint may serve as an example. At the macroscopic level the particle distinguishability manifests itself as the additional nonlinear flux which is responsible for the mixing of different species of a lattice gas. This flux leads to the array of anomalous diffusive transport effects, e.g., the drag effect, formation of the drifting spatial structures \cite{schmittmann_statistical_1995,leung_drifting_1997,hipolito_effects_2003}, effects of ``nega\-tive'' mass transport \cite{lukyanets_drift_2010,N11Chum}.

When the difference between the values of the component mobility is significant, the nonlinear flux can lead to indirect correlations in the slow subsystem. These correlations are induced by the light component that plays the role of interaction mediator for the particles of the slow subsystem. Such a phenomenon is similar to the hydrodynamic interactions between colloidal particles \cite{ColloidBook,alder_decay_1970,pagonab_short-time_1999}.
One of the intriguing effects observed in these suspensions is the presence of long-time correlations between liquid-embedded particles when the velocity correlation function of a particle pair is characterized by the long-time negative tail \cite{hagen_1999,frydel_long-range_2010,felderhof_long-range_2011}.

In this paper we are aimed to show the possibility of similar long-time correlations for the slow component of a lattice gas. For this purpose, we consider the simplest case of a two-component lattice gas by using the adiabatic and the uniform clusters approximations.

\section{Correlations in a two-component lattice gas}
\label{sec:1}
We consider the simplest two-component lattice gas model subject to an excluded volume constraint. The kinetics is defined by the jumps of particles to the neighboring vacant sites. The variation of the $i$-th site occupancy by the particles of sort $\alpha$ during the time interval $\Delta t$, $\tau_0\ll\Delta t\ll\tau_l$ ($\tau_0$ is the duration of a particle jump to a neighboring site, $\tau_l$ being the life time of a particle on a site), is described by the standard continuity equation (see, e.g., \cite{chumak_diffusion_1980,tahir-kheli_correlated_1983})
\begin{equation}\label{balance+}
n_i^\alpha (t+\Delta t)-n_i^\alpha(t)=\sum_j \left(J^\alpha_{ji}-J^\alpha_{ij}\right) + \delta J_i^\alpha,
\end{equation}
where $n_i^\alpha=0,1$ are the local occupation numbers of particles $\alpha$ at the $i$-th site, $J^\alpha_{ij}=\nu^\alpha_{ij} n_i^\alpha h_j \Delta t$ gives the mean number of jumps (of particles $\alpha$ from site $i$ to a neighboring site $j$ per time $\Delta t$), $\nu_{ij}^\alpha=\nu_\alpha$ is the frequency of these jumps, $h_j=1-\sum_\beta n_j^\beta$ is the occupation number of the $j$th site vacancies. The term $\delta J_i^\alpha$ stands for the Langevin source of the $i$th site occupation number fluctuations. Equations for the average local occupation numbers may be obtained from Eqs.~(\ref{balance+}), using the local equilibrium approximation (Zubarev approach) \cite{chumak_diffusion_1980,zubarev_nonequilibrium_1974}, which in our case coincides with the mean field approximation \cite{leung_novel_1994}. In the case of two components, introducing time derivatives in Eqs.~(\ref{balance+}), see \cite{richards_theory_1977}, the macroscopic equations take the form \cite{schmittmann_statistical_1995,hipolito_effects_2003,lukyanets_drift_2010}
\begin{subequations}\label{3}
\begin{equation}
\nu_m^{-1} \dot m = \nabla^2m + \nabla (m\nabla n - n\nabla m)+\delta\tilde J_m, \label{3a}
\end{equation}
\begin{equation}
\nu_n^{-1} \dot n = \nabla^2n - \nabla (m\nabla n - n\nabla m)+\delta\tilde J_n, \label{3b}
\end{equation}
\end{subequations}
where $m$ and $n$ are the average occupation numbers of the two components at the point $r$, $\nu_m$ and $\nu_n$ are the jump frequencies of particles $m$ and $n$, correspondingly. These equations are obtained in the long-wavelength approximation assuming that the characteristic scale of the gas inhomogeneity is much grater than the lattice constant $a$. The dimensionless spatial coordinate $\mathbf{r}\rightarrow \mathbf{r}/a$ is introduced.

Distinguishability of the particles in the noninteracting lattice gas leads to the additional nonlinear flow $n\nabla m - m \nabla n$ in the macroscopic Eqs.~(\ref{3}) that is absent in the case of a one-component gas. An important condition for the onset of this additional term is the excluded volume constraint, i.e., when a site may be occupied by one particle only. This flow is responsible for the mixing of different particle species and leads to additional correlations in the system.

In the case of considerable difference between the species mobilities, $\nu_n/\nu_m\!\ll\!1$, the indirect correlations can occur in the slow subsystem $n$. These correlations are induced by the fast component that plays the role of the ``interaction mediator''. Such correlations are similar to the hydrodynamic correlations between colloidal particles \cite{frydel_long-range_2010} and may have the long-time character.

In order to show the presence of the long-time correlations in the heavy subsystem of the lattice gas we consider the fluctuations of the heavy particles drift caused by an external field. They manifest itself as the fluctuations of hydrodynamic velocity field $\mathbf{u}(\mathbf{r},t)$. In what follows, we assume that the characteristic time-scale $\tau_u$ of the fluctuations $\mathbf{u}(\mathbf{r},t)$ satisfies the condition $\nu_m\gg\tau_u^{-1}\gg\nu_n$ and restrict ourselves to the adiabatic approximation, $\nu_m\rightarrow \infty$. It means that the distribution of the fast component instantly gets into the stationary one with the momentary distribution of the slow component. Then, kinetic equations take form
\begin{subequations}\label{adiabat}
\begin{alignat}{2}
0&= \nabla^2m & &+ \nabla (m\nabla n - n\nabla m), \label{adiabata}\\
\nu_n^{-1} \dot n &= \nabla^2n & &- \nabla (m\nabla n - n\nabla m)+\nu_n^{-1}u(\mathbf{r},\tau)\nabla n, \label{adiabatb}
\end{alignat}
\end{subequations}
The light component $m(r,t)$ depends on a ``slow'' time $t$ as on a parameter. It should be noted, that the fluctuations of drift are associated with the fluctuations of the particle jump probability $\nu_{i,i+1}^n=\nu_n+u_i$ between the neighboring lattice sites $i$, $i+1$ and in general case this term may have a more complicated form. For example, in the case of the asymmetric jumps of the lattice gas particles, $\nu_{i,i+1}^n=\nu_n+u$, $\nu_{i,i-1}^n=\nu_n$, the drift term reads $\nu_n^{-1}u(\nabla n-\nabla n^2-\nabla nm)$. However, here we restrict ourselves to the case given by Eqs.~(\ref{adiabat}) and, similarly to \cite{frydel_long-range_2010}, consider a quasi-one-dimensional situation when interparticle correlations become most significant. In this case, one can easily eliminate the fast variable $m$ from Eqs.~(\ref{adiabat}), see \cite{lukyanets_drift_2010}, and obtain the self-consistent equation for the heavy component $n$
\begin{equation}
\frac{\partial n(x,\tau)}{\partial \tau}
=\left(1-m_1+J(\tau)\int_{-L}^x\frac{\mathrm{d}s}{[1-n(s,\tau)]^2}\right)\frac{\partial^2n(x,\tau)}{\partial x^2} +\nu_n^{-1}u(x,\tau)\frac{\partial n(x,\tau)}{\partial x}, \label{Nonlocal}
\end{equation}
\begin{equation}
J(\tau)=-(m_2-m_1)\left(\int_{-L}^L\frac{\mathrm{d}s}{[1-n(s,\tau)]^2}\right)^{-1}. \label{J}
\end{equation}
Here we introduced the dimensionless time $\tau\!=\!\nu_nt$. The quantity $J(\tau)$ [Eq.~(\ref{J})] corresponds to the average macroscopic flow of the light component $m(x,\tau)$ passing through the system \cite{lukyanets_drift_2010}; $m(-L,\tau)\!=\!m_1\!=\!\mathrm{const}$, $m(L,\tau)\!=\!m_2\!=\!\mathrm{const}$, and $n(-L,\tau)\!=\!n(L,\tau)\!=\!0$ being the values of the mean occupation numbers $m$ and $n$ at the system boundaries $x=\mp L$.

It follows from Eq.~(\ref{Nonlocal}) that the correlations in the slow subsystem appear only in the nonequilibrium case with the nonzero flow $J(\tau)$ that is responsible for the nonlinear and nonlocal character of the diffusion process.

Next we make use of the uniform density cluster approximation that is often exploited, e.g., for the description of Coulomb explosions of deuterium clusters \cite{zweiback_nuclear_2000,li_coulomb_2006}. Under this approximation, the density distribution $n(x,\tau)$ is considered as a system of uniform one-dimensional clusters. It is justified when the cluster size is much smaller than that of the system. Each such cluster is characterized by its ``center of mass'' $r_k(\tau)$ and width $R_k(\tau)$, so that the distribution of the slow component can be written as
\begin{equation} \label{napprox}
n(x,\tau)=\left\{
\begin{array}{ll}
n_k(\tau)=Q_k/R_k(\tau) & \textrm{if } |x-r_k(\tau)|\leq R_k(\tau)/2\textrm{,}\\
0 & \textrm{if } |x-r_k(\tau)|>R_k(\tau)/2\textrm{,}
\end{array} \right.
\end{equation}
where $Q_k$ is the number of the particles contained by the $k$th cluster. Further, we assume the characteristic spatial scale $\xi$ of the fluctuations $u(x,\tau)$ to be much grater than the typical cluster size but less than the distance between them, $R_k\ll\xi\ll|r_{k+1}-r_k|$. Then, evaluating the moments of distribution (\ref{napprox}) from Eq.~(\ref{Nonlocal}) one can obtain the equations for $r_k(\tau)$ and $R_k(\tau)$
\begin{subequations}\label{apro2}
\begin{equation}\label{apro2_a}
\frac{\partial r_k(\tau)}{\partial \tau} = J(t) G[n_k(\tau)]-\nu_n^{-1}u_k(\tau),
\end{equation}
\begin{equation}\label{apro2_c}
\frac{1}{12}\frac{\partial R_k^2(\tau)}{\partial \tau}=2D[r_k(\tau)]+J(\tau)\left(\sum_{l<k}Q_lG[n_l(\tau)]-\sum_{l>k}Q_lG[n_l(\tau)]\right).
\end{equation}
\end{subequations}
The first equation (\ref{apro2_a}) describes the drift velocity $\dot r_k$ of the $k$th cluster. The motion of the cluster as a whole is induced by the fluctuations of the external field $u_k$ and by the flow $J(\tau)\!=\!2Lj\{2L\!+\!\sum_kQ_kG[n_k(\tau)]\}^{-1}$ of the fast component (the drag effect). The function $G[n_k(\tau)]\!=\![1-n_k(\tau)]^{-1}+[1-n_k(\tau)]^{-2}$ describes the drag velocity slowdown caused by the cluster expansion. The second equation (\ref{apro2_c}) describes the rate of the cluster expansion that is defined by the diffusion coefficient $D[r_k(\tau)]=1-(m_1+m_2)/2+J(\tau)r_k(\tau)$ to be cluster position dependent and by the drag velocities $JG(n_k)$ of other clusters, see (\ref{apro2_a}).
Note that the quantity $j\!=\!-\Delta m/(2L)\ll1$ is a natural small parameter since $|\Delta m|\!=\!|m_2\!-\!m_1|\!\leq\!1$.

In fact Eqs.~(\ref{Nonlocal}) describe the Brownian-like dynamics of the penetrable, expanding clusters embedded into the flow of the fast component.
For simplicity, next we consider the system of the two initially identical clusters, $Q=Q_{1(2)}$, $R(0)=R_{1(2)}(0)$, with the $\delta$-correlated fluctuations, $\langle u_k(\tau)u_l(\tau')\rangle=\Gamma\delta_{kl}\delta(\tau-\tau')$.

We are interested in the response of one cluster caused by the initial velocity perturbation of another cluster.
For this purpose, we consider the pair correlation function of velocity fluctuations $C_{12}(\tau)=\langle \delta v_1(\tau)\delta v_2(0)\rangle = \langle [\dot r_1(\tau)-\bar v_1(\tau)][\dot r_2(0)-\bar v_2(0)]\rangle$,
$\bar v_k(\tau)=\dot{\bar{r}}_k(\tau)$, where $\bar{r}_k(\tau)$ satisfies unperturbed Eqs.~(\ref{apro2}) (i.e., Eqs.~(\ref{apro2}) provided $u_k(\tau)\!\equiv\!0$).
Supposing that the fluctuations $u_k(t)$ are small, $\nu_n^{-1}|u_k| \ll |\bar v_k(t)|$, and using the perturbation theory in $\nu_n^{-1}|u_k|$ and $j$ (we assume, that $j^2\ll\nu_n^{-1}|u_k|\ll j$), one can get
\begin{equation}\label{Correlator_full}
C_{12}(\tau)\approx\nu_n^{-2} j^3\frac{24\Gamma Q}{D_0}\left(\frac{R(\tau)}{[R(\tau)-Q]^4}+\frac{2QR(\tau)}{[R(\tau)-Q]^5}-\frac{C_0R(\tau)}{[R(\tau)-Q]^3}\right),
\end{equation}
where $R(\tau)\!=\![24D\tau+R^2(0)]^{1/2}$, $D_0=1-(m_1+m_2)/2$ and  $C_0=[R(0)+Q]/[R(0)-Q]^2$.

\begin{figure}
\includegraphics[width=0.45\columnwidth]{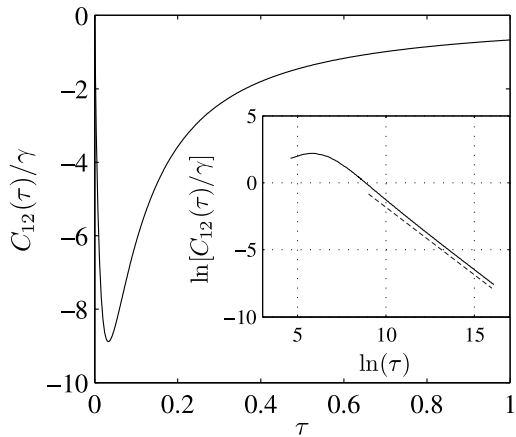}
\caption{\label{fig:1}
Velocity cross-correlation function of a cluster pair as a function of time [$Q=0.5$, $R(0)=1$, $\Delta m =1$, $\gamma=24\Gamma Q j^3/(D_0\nu_n^2)$]. The inset shows the long-time asymptotic behavior, the dashed line indicates a slope of 1.}
\end{figure}

The correlation function (\ref{Correlator_full}), Fig.~(\ref{fig:1}),
has the long-time negative tail with the asymptotic behavior
  $C_{12}(\tau\rightarrow\infty)\simeq -\tau^{-1}$. 
The long-time correlations between the two diffusive clusters are similar to hydrodynamic ones appearing between colloidal particles confined in a liquid-filled linear channel \cite{frydel_long-range_2010}.
\section{Summary and discussion}
\label{sec:3}
The significant difference of particle mobilities in a two-component lattice gas leads to the induced correlations in the slow subsystem mediated by the fast one. They manifest itself as a negative long-time tail in the behavior of the pair correlation function and are similar to the well-known long-time correlations between colloidal particles \cite{frydel_long-range_2010,felderhof_long-range_2011}. Note that there is no additional interparticle interaction introduced in the lattice gas except that due to hard-core exclusion. In order to demonstrate this phenomenon we have considered the one-dimensional case and used the simplest, somewhat rough, approximations. In particular, we have used the adiabatic one and neglected the retardation effects. As a result, the correlations instantly spread all over the system and the averaged flow of the fast component does not depend on the distance between the clusters of the slow one. It give rise to the weak dependence of the correlation function on the inter-cluster distance that make itself evident in the higher orders of the perturbation theory expansion.

Note that the correlations in two- and three-dimensional cases may be considered in the same manner by using the uniform density cluster approximation.

Considered correlations may be  associated with the ``elastic-type interactions'', similar to the case of the liquid-embedded particles \cite{lamb_hydrodynamics_2012,pooley_hydrodynamic_2007}. However, another type of the induced correlations in the slow component may be caused by the equilibrium fluctuations in the fast one and associated with so-called Casimir-like forces \cite{bitbol_forces_2011}, that is beyond the scope of this paper.

\end{document}